\documentclass[conference]{IEEEtran}
\IEEEoverridecommandlockouts
\usepackage{footnote}
\makesavenoteenv{tabular}
\usepackage{cite}
\usepackage{amsmath,amssymb,amsfonts}
\usepackage{algorithmic}
\usepackage{graphicx}
\usepackage{textcomp}
\usepackage{booktabs}
\usepackage{xcolor}
\def\BibTeX{{\rm B\kern-.05em{\sc i\kern-.025em b}\kern-.08em
    T\kern-.1667em\lower.7ex\hbox{E}\kern-.125emX}}

\usepackage{siunitx}
\usepackage{xcolor}
\usepackage{hyperref}

\begin{document}

\title{BUT System Description for DIHARD \\ Speech Diarization Challenge 2019}

\author{\IEEEauthorblockN{Federico Landini$^1$,  Shuai Wang$^{1,2}$, Mireia Diez$^1$, Luk\'{a}\v{s} Burget$^1$, Pavel Mat\v{e}jka$^1$, Kate\v{r}ina \v{Z}mol\'{i}kov\'{a}$^1$, \\Ladislav Mo\v{s}ner$^1$, Old\v{r}ich Plchot$^1$, Ond\v{r}ej Novotn\'{y}$^1$, Hossein Zeinali$^1$, Johan Rohdin$^1$}

    \IEEEauthorblockA{
        $^1$Brno University of Technology, Speech@FIT, Czechia \\
        $^2$ SpeechLab, Shanghai Jiao Tong University, China \\
        \textit{\{landini,mireia\}@fit.vutbr.cz}
    }
}

\maketitle

\begin{abstract}
This paper describes the systems developed by the BUT team for the four tracks of the second DIHARD speech diarization challenge \cite{ryant2019second}. For tracks 1 and 2  \cite{bergelson2016bergelson, ryant2019dihard} the systems were based on performing agglomerative hierarchical clustering (AHC) over x-vectors, followed by the Bayesian Hidden Markov Model (HMM) with eigenvoice priors applied at x-vector level followed by the same approach applied at frame level. For tracks 3 and 4 \cite{barker2018fifth}, the systems were based on performing AHC using x-vectors extracted on all channels.
\end{abstract}

\begin{IEEEkeywords}
Speaker Diarization, Variational Bayes, HMM, x-vector, WPE, VAD, Overlapped Speech, DIHARD, CHiME
\end{IEEEkeywords}

\section{Track 1}
\label{sec:track1}

\subsection{System 1}
\label{subsec:system1}

\begin{table}[h]
\centering
\begin{tabular}{c|c|c} 
Performance on  & DER [\%]  & JER\\ 
\midrule
development\footnotemark & 18.09 & 42.81\\
evaluation  & 18.42 & 44.58\\
\end{tabular}
\end{table}
\footnotetext{Note that the performance on the development set is overoptimistic for this system as it uses a PLDA model supervisedly adapted on the same data.}

Diarization of each recording consists of the following steps: x-vectors~\cite{SnyderIS17} are extracted from speech sub-segments (as given by the reference Voice Activity Detection (VAD)) every 0.25s from windows of 1.5s. These x-vectors are clustered using AHC with similarity metric based on Probabilistic Linear Discriminant Analysis (PLDA)~\cite{kenny10PLDA_HTP} log-likelihood ratio scores as used for speaker verification. The threshold used as stopping criterion for clustering is set to under-cluster (i.e. to result with more speaker clusters) as this step is meant only as an initialization for the following more precise clustering step based on Bayesian Hidden Markov Model (BHMM). Our submission to the previous DIHARD challenge~\cite{diez2018but} used similar BHMM with Mel-Frequency Cepstral Coefficients (MFCC) features as input. However, our current system first uses BHMM with x-vectors at the input. In~\cite{DiezInter19}, we have shown that BHMM provides significantly better results as compared to the simple AHC. Still it is beneficial to initialize the iterative Variational Bayes (VB) inference in the BHMM from the output of the previous AHC step. Since the x-vectors are extracted with relatively low resolution in time, we apply a re-segmentation step to refine the speaker change boundaries. For this purpose, we use the BHMM with MFCC features~\cite{DiezOdyssey18,Dieztobe} as input. The frame-to-speaker assignments are initialized from the previous step and only a single iteration of the VB algorithm is used (to update speaker models and) to obtain the final frame-to-speaker assignments. Finally, we apply a post-processing step which tries to detect and deal with overlapped speech. In the following paragraphs, more details are provided on the individual steps and how to train the corresponding models.

\textbf{Signal processing}\label{dereverb-WPE}
We used the Weighted Prediction Error (WPE) \cite{nakatani-wpe-2010,Drude2018NaraWPE} method to remove late reverberation from the data. We estimated a dereverberation filter on Short Time Fourier Transform (STFT) spectrum for every 100 seconds block of an utterance. To compute the STFT, we used \SI{32}{\milli\second} windows with \SI{8}{\milli\second} shift. We set the filter length and prediction delay to 30 and 3 respectively for \SI{16}{\kHz}. The number of iterations was set to 3.

\textbf{X-vector extractor} \label{xvecs}
The systems were trained with the Kaldi toolkit~\cite{povey2011kaldi} using the SRE16 recipe~\cite{snyder_kaldi_recipe} with the modifications described below:
\begin{itemize}
    \item Feature sets consist of 40 FBANKs with 16kHz sampling frequency.
    \item Networks trained with 6 epochs (instead of 3).
    \item Using a modified sample generation - we used 200 frames in all training segments instead of randomizing it between 200-400 frames. With the random process used for generating segments in the original Kaldi recipe, some recordings can be under-represented (too few or none samples chosen from them) and other over-represented (too many segments chosen from them). In order to avoid this, we sample from all the recordings considering all possible segments in order to be sure that all of them are used.
    \item The x-vector Deep Neural Network (DNN) was trained on VoxCeleb 1 and 2 \cite{chung2018voxceleb2} with 1.2 million speech segments from 7146 speakers plus additional 5 million segments obtained with data augmentation.
    We generated around 700 Kaldi archives such that each of them contained exactly 15 training samples from each speaker (i.e. around 107K samples in each archive).
    \item The architecture of the network for x-vector extraction is shown in Table~\ref{tab:xv_topology}.
\end{itemize}

\begin{table*}[tb]
\renewcommand{\arraystretch}{0.85}
\centering
\caption{\label{tab:xv_topology}x-vector topology proposed in \cite{snyder2019speaker}. K in the first layer indicates different feature dimensionalities, T is the number of training segment frames and N in the last row is the number of speakers.}
\begin{tabular}{c|c|c} \toprule
 Layer & \textbf{Layer context}  & \textbf{(Input) $\times$ output}\\ 
 \midrule
 frame1 & $[t-2, t-1, t, t+1, t+2]$    & (5 $\times$ K) $\times$ 1024 \\
frame2 & $[t]$                        & 1024 $\times$ 1024\\
frame3 & $[t-4, t-2, t, t+2, t+4]$    & (5 $\times$ 1024) $\times$ 1024\\
frame4 & $[t]$                        & 1024 $\times$ 1024 \\
frame5 & $[t-3, t, t+3]$              & (3 $\times$ 1024) $\times$ 1024\\
frame6 & $[t]$                        & 1024 $\times$ 1024  \\
frame7 & $[t-4, t, t+4]$              & (3 $\times$ 1024) $\times$ 1024\\
frame8 & $[t]$                        & 1024 $\times$ 1024  \\
frame9 & $[t]$                        & 1024 $\times$ 2000 \\
stats pooling (frame7,frame9) & $[0, T]$    & 2*1024+2*2000 $\times$ 512 \\
segment1 & $[0, T]$                      & 512 $\times$ 512\\
softmax  & $[0, T]$ & 512 $\times$ N \\
\bottomrule
\end{tabular}
\end{table*}

\textbf{PLDA estimation}
The PLDA model is used to calculate the similarity scores for AHC. The same PLDA is also used to model speaker distributions in the x-vector based BHMM as described in detail in~\cite{DiezInter19}. The PLDA model is trained on x-vectors extracted from 3s speech segments from VoxCeleb 1 and 2 and utterance IDs combined with speaker IDs serve as the class labels. Before the PLDA training, the x-vectors are centered (i.e. mean normalized), whitened (i.e. normalized to have identity covariance matrix) and length-normalized~\cite{GarciaRomero2011lnorm}. The centering and whitening transformation are estimated on the joint set of DIHARD development and evaluation data.

To better take advantage of the in-domain data, another PLDA model is trained on x-vectors extracted in a similar way using the DIHARD development data. In this case, the centering and whitening transformation are also estimated on the joint set of DIHARD development and evaluation data. Note that these transformations are also applied to the evaluation x-vectors when performing diarization.

The adapted PLDA model used for both AHC and BHMM based clustering is obtained as an interpolation of the two PLDA models: means, within- and across-class covariance matrices from the two models are averaged (using equal weights).

\textbf{i-vector extractor}
Although we do not use any i-vectors in our diarization system, the MFCC based BHMM uses eigenvoice priors~\cite{DiezOdyssey18,KennyBayDiar} to robustly model speaker distributions, which is simply equivalent to an i-vector extractor model. In our system, the i-vector extractor is based on a diagonal Universal Background Model composed of a Gaussian Mixture Model (UBM-GMM) with 1024 Gaussian components, 400-eigenvoices and 40-dimensional features (19 MFCC + Energy + deltas) extracted from 16kHz speech, 25ms window, 10ms frame rate, no cepstral mean normalization. It is trained on VoxCeleb 2 data.

\textbf{AHC clustering}\label{initial-clustering}
The clustering is essentially the same as implemented in the official Kaldi diarization recipe~\cite{sell2018diarization}: For each test recording, Principal Component Analysis (PCA) is estimated on centered, whitened and length normalized x-vectors from that recording. Using PCA, the x-vectors are projected into a low dimensional space preserving 35\% of their variability\footnote{The original Kaldi recipe preserves only 10\% variability, which corresponds to about 5 dimensions. In our case, 20 dimensions are preserved in average, which we have found to significantly improve the performance on DIHARD data.} and the resulting x-vectors are again length normalized. The PLDA parameters are also projected to the corresponding low-dimensional space. The resulting PLDA model is used to calculate log-likelihood ratio verification scores as a similarity metric for each pair of x-vectors from the test recording. The resulting pair-wise similarity matrix is the only input to the {\em unweighted average linkage} AHC (i.e. UPGMA method is used).

The similarity score threshold for stopping the AHC process is estimated for each recording separately using an {\em unsupervised linear calibration}: a GMM with two univariate Gaussian components with shared variance is trained on all the scores from the similarity matrix. The two Gaussian components are assumed to be the score distributions corresponding to the same-speaker and different-speaker x-vector pairs. Therefore, the threshold is set as the score for which the posterior probability of both components is 0.5 (i.e. decision threshold for the same/different-speaker maximum-a-posteriori classifier). To encourage under-clustering at this step, we add a constant bias to this threshold. However, $threshold\_bias=0.0$ was found optimal in the case of this system\footnote{Note that other systems use non-zero values for this parameter. }.

\textbf{Variational Bayes HMM at x-vector level}\label{VB-xvector}
BHMM is used to cluster x-vectors as described in detail in~\cite{DiezInter19}. Each iteration of BHMM refines the probabilistic assignment of x-vectors to speaker clusters. The initial assignment is taken from the previous AHC step. AHC provides hard assignments of x-vectors to clusters, which we further ``smooth out'' as follows: for each x-vector, its cluster label provided by AHC is represented as a one-hot-encoding vector. This vector is multiplied by a {\em smoothing factor} (5.0 for this system) and re-normalized by the softmax function to give the initial soft (probabilistic) assignment of the x-vector to speaker clusters.

Before the BHMM clustering, the x-vectors are projected using Latent Discriminant Analysis (LDA) into a 220 dimensional space. The LDA projection is calculated directly from the parameters of the PLDA model described above. The PLDA model itself is also projected into the 220-dimensional space and used in the BHMM to model speaker distributions as described in~\cite{DiezInter19}.

Iterative VB inference is run until convergence to update the assignment of x-vectors to speaker clusters. Automatic Relevance Determination (ARD)~\cite{Bishop2006} inherent in BHMM results in dropping the redundant speaker clusters and allows us to properly estimate the number of speakers in each recording. To optimize the diarization performance, VB inference in the BHMM model is controlled by a number of parameters, which are set to the indicated values for this system: in~\cite{DiezInter19}, we have newly introduced the \textit{speaker regularization coefficient $F_B=11.0$}, which controls ARD to be more or less aggressive when dropping the redundant speakers (i.e. tune the model to find the right number of speaker clusters). \textit{Acoustic scaling factor $F_A=0.4$} is introduced to compensate for the incorrect assumption of statistical independence between observations (i.e. x-vectors). $P_\text{loop}=0.8$ is the probability of not changing speakers when moving to the next observation, which serves as speaker turn duration model. Note again that our system uses a relatively high frame rate of 4 x-vectors per second which asks for the high value of $P_\text{loop}$ and low value of $F_A$ as compared to the optimal values reported in~\cite{DiezInter19}.

\textbf{Variational Bayes HMM at frame level}\label{VB-per-frame} Given that the output from the previous clustering step has a time resolution of 0.25s, we apply a second step of the BHMM clustering at MFCC features frame level. The algorithm is applied in a similar manner as in \cite{diez2018but} using the diarization output produced by the previous step as initialization and MFCCs as frame features. However, instead of running the algorithm until convergence, only one iteration is performed in order to only correct boundaries between speakers. The speaker subspace corresponds to the i-vector extractor described above.
The parameters used for the VB algorithm~\cite{burgetVB} were: $F_A=0.1$, $P_\text{loop}=0.95$, min duration 1 and downsampling 5.

\textbf{Overlap post-processing}\label{overlap} Given that all the models up to this point assume that each representation of speech (either x-vectors or per-frame features) have speech of only one speaker, we perform overlapped speech detection and apply a heuristic to label segments for more than one speaker. Silence segments are removed from the recording and x-vectors are extracted every 0.25s from windows of 1.5s on the concatenated segments of speech. 
The x-vectors are labeled as overlapped speech if more than half of the original segment contains overlapped speech and as non-overlapped speech otherwise. 
Then, a logistic regression model was trained to predict overlapped/non-overlapped speech segments. 
The threshold used for detection was 0.7. Once overlap segments are detected, the heuristic consists in assigning for each frame in an overlapped speech segment the two closest speakers according to the diarization labels given by the previous step.

\subsection{McClane}
\label{subsec:mcclane}

\begin{table}[h]
\centering
\begin{tabular}{c|c|c} 
Performance on  & DER [\%]  & JER\\ 
\midrule
development & 18.06 & 44.04\\
evaluation  & 18.74 & 45.59\\
\end{tabular}
\end{table}

This system is essentially the same as System 1 described in Section~\ref{subsec:system1} except that  the PLDA model is trained only on Voxceleb data and it is not adapted or trained on any DIHARD data. Also, some of the system parameters were tuned to different values. Specifically, for AHC, PCA was set to preserve $22\%$ of variability and $threshold\_bias=0.2$. BHMM clustering of x-vectors uses LDA dimensionality 250, $P_\text{loop}=0.95$ and $F_B=12.0$.

\section{Track 2}
\label{sec:track2}

\subsection{System 2}
\label{subsec:system2}

\begin{table}[h]
\centering
\begin{tabular}{c|c|c} 
Performance on  & DER [\%]  & JER\\ 
\midrule
development & 23.94 & 46.67 \\
evaluation  & 27.26 & 49.15 \\
\end{tabular}
\end{table}

This system is essentially the same as System 1 described in Section~\ref{subsec:system1} except that the PLDA model is trained only on Voxceleb data and it is not adapted or trained on any DIHARD data. Also, some of the system parameters were tuned to different values. Specifically, for AHC, PCA was set to preserve $30\%$ of variability and $threshold\_bias=0.2$. BHMM clustering of x-vectors uses LDA dimensionality 250, $P_\text{loop}=0.7$ and $F_B=12.0$. Threshold of 0.8 is used for the overlap detection post processing. Most importantly, the reference VAD is not available for the evaluation data in track 2. Therefore, in all processing steps involving the evaluation data, the reference VAD was replaced by an automatic VAD described in the following paragraph.  

\textbf{Voice Activity Detection} A DNN-based system was trained for binary,  speech/non-speech, classification of speech frames using the development set as training data. The network consists of 3 hidden layers of 512 nodes each and frame features are 40 dimensional FBANKs extracted with a 25ms frame window width and 10ms. The input to the network is obtained by stacking a frame with the 5 preceding and 5 subsequent frames leading to a 440 dimensional input. After obtaining the frame-level prediction, the output is smoothed by means of a median filter with a 25-frame window size.

\subsection{System 2b}
\label{subsec:system2b}

\begin{table}[h]
\centering
\begin{tabular}{c|c|c} 
Performance on  & DER [\%]  & JER\\ 
\midrule
development & 23.81 & 46.64 \\
evaluation  & 27.11 & 49.07\\
\end{tabular}
\end{table}

This system is the same as System 2 described in Section~\ref{subsec:system2} except that it does not use the overlap detection post processing.

\section{Track 3}
\label{sec:track3}

\subsection{System 3}
\label{subsec:system3}

\begin{table}[h]
\centering
\begin{tabular}{c|c|c} 
Performance on & DER [\%]  & JER\\ 
\midrule
dev+train & 53.81 & 60.73 \\
evaluation  & 45.65 & 60.12 \\
\end{tabular}
\end{table}

\textbf{Signal pre-processing} The WPE method is used in the same fashion as described in \ref{dereverb-WPE} applying it independently to each of the four channels.

\textbf{X-vector extractor} The x-vectors are extracted in the same fashion as explained in \ref{xvecs} independently on each of the four channels.

\textbf{Clustering}\label{clustering} The initial clustering is performed in a similar manner as explained in \ref{initial-clustering}. In this case, the PLDA model is adapted in an unsupervised way to the train and development data from the CHiME corpus. Given that four channels per recording device are available, the same procedure is carried out in each one individually to produce the score matrices for each channel. The four matrices are averaged with equal weights before doing the clustering and the clustering is performed until convergence using a threshold of 2.1.

\section{Track 4}
\label{sec:track4}

\subsection{System 4}
\label{subsec:system4}

\begin{table}[h]
\centering
\begin{tabular}{c|c|c} 
Performance on & DER [\%]  & JER\\ 
\midrule
dev+train & 66.86 & 68.19 \\
evaluation  & 58.92 & 65.08 \\
\end{tabular}
\end{table}

\textbf{Signal pre-processing} The WPE method is used in the same fashion as described in \ref{dereverb-WPE} applying it independently to each of the four channels.

\textbf{Voice Activity Detection} Given that for track 4 the oracle VAD labels are not available, a VAD system is used in order to discard silence and feed the rest of the system only with speech segments. The VAD labels were obtained by running the system on the recordings of Channel 1 downsampled to 8kHz. 
The VAD system used is based on a neural network (NN) trained for binary, speech/non-speech, classification of speech frames. The 288-dimensional NN  input is derived from 31 frames of 15 log Mel filter-bank outputs and 3 pitch features.
The NN with 2 hidden layers of 400 sigmoid neurons was trained on the Fisher English dataset \cite{cieri2004fisher} with labels provided from automatic speech recognition alignment. Per-frame logit posterior probabilities of speech are smoothed by averaging over consecutive 31 frames and thresholded to at the value of 0 to give the final hard per frame speech/non-speech decision. See~\cite{Matejka:LRE17} for a more detailed description of the VAD system.

\textbf{X-vector extractor} The x-vectors were extracted in the same fashion as explained in \ref{xvecs} with the difference that we used only the speech segments defined by the VAD to compute them.

\textbf{Clustering} The clustering was performed in the same way as explained in \ref{clustering} with the difference that the threshold used as stopping criterion was 1.8.

\section{Hardware requirements}

The infrastructure used to run the experiments was a CPU, Intel(R) Xeon(R) CPU 5675 @ 3.07GHz, with a total memory of 37GB unless specified otherwise.

For WPE dereverberation, the processing time for 1 minute of audio is approximately 6s for single-channel and 15s for multi-channel data. 

The computation of the VAD labels used for Track 2 is performed on GPU (Tesla P100 PCIe 16GB), the processing time for 1 minute of audio is approximately 0.067s.

For the VAD labels used for the Track 4, the processing time for 1 minute of audio is approximately 1.5s for each channel.

The processing time for the x-vector extraction of 1 minute of audio is 4s.

AHC has the time complexity $\mathcal{O}(n^3)$ and memory complexity $\mathcal{O}(n^2)$, which is in both cases the highest (theoretical) complexity out of all the processing steps. Therefore, AHC could become the bottleneck for very long utterances. However, even for the longest utterances in the DIHARD evaluation data (around  10 minutes), our non-optimized python implementation of AHC is still about 20 faster than realtime. On average, on all the DIHARD evaluation recordings (taking into account only speech and not silence), AHC is around 100 faster than real-time.

BHMM based clustering of x-vectors initialized from the AHC is more than 200 times faster than real-time and the MFCC based BHMM clustering is more than 30 times faster than real-time on average on the DIHARD evaluation recordings.

Post-processing the speaker labels for overlap detection on a recording of 10 minutes varied from less than a second to 1 minute depending on the amount of overlapped speech segments found.

\bibliographystyle{IEEEtran}
\bibliography{mybib,biblio}

\end{document}